# A Paradigm Shift: The Implications of Working Memory Limits for Physics and Chemistry Instruction

JudithAnn R. Hartman,* Eric A. Nelson†

*Department of Chemistry (Emerita), United States Naval Academy, Annapolis, MD 21042

†Fairfax County Public Schools (Retired), Fairfax, VA. 22030

**ABSTRACT**

Scientists who study how the brain solves problems have recently verified that, because of stringent limitations in "working memory" (which is where the brain solves problems), to reliably solve problems of any complexity, students must apply facts and algorithms that have previously been well memorized. This is a paradigm shift: A change in science's fundamental understanding of how the brain solves problems and how we can best guide students to learn to solve problems in the physical sciences. One implication is that for students, knowledge of concepts and "big ideas" is not sufficient to solve most problems assigned in physics and chemistry courses for STEM majors. To develop an intuitive sense of which fundamentals to recall when, first students must make the fundamental relationships of a topic "recallable with automaticity" then apply those fundamentals to solving problems in a variety of distinctive contexts. Based on these findings, cognitive science has identified strategies that speed learning and assist in retention of physics and chemistry. Experiments will be suggested by which instructors can test science-informed methodologies.

**INTRODUCTION**

Instruction in physics and chemistry is informed by two sciences: the science of the discipline and cognitive science.

Cognitive science is the study of how the brain thinks, conducted by scientists from disciplines including neuroscience, cognitive and educational psychology, and evolutionary biology. A sub-discipline studies how the brain solves problems and learns to solve problems. For educators in the physical sciences, guiding students in learning to solve problems is a central part of our work. Knowing what science has found the brain can and cannot manage during problem solving assists in designing effective instruction.

In the past two decades, with assistance from new technologies,[1-3] cognitive experts have reached consensus on a number of principles that describe how the brain solves problems in chemistry, physics, and mathematics.[4,5] In 2017, neuroscientist Mark Seidenberg summarized, "Although we are far from understanding how the brain works, many of its fundamental properties are known."[6] These fundamentals hold the promise of transforming the basis of STEM education from opinion-based theories to science-informed best practices.

For a number of years, the authors of this paper have been involved in an informal collaboration of higher education and Advanced Placement (AP) instructors seeking to apply cognitive science to improve achievement in introductory physical science courses. In previously published papers, those instructors have reported measured positive results.[7,8,9] Based on that experience, we believe that by applying findings of cognitive research, instructors can help students learn chemistry and physics faster and with better retention.

However, even among scientists, the progress of science can be controversial.[10] When experiments yield unexpected findings, difficult questions must be asked about assumptions that have guided previous scientific work.

Prior to 1990, substantial memorization was accepted as a necessary part of learning physics and chemistry. But with new technologies, as information became increasingly accessible, views on memorization changed for many educators. Reforms were proposed and often adopted based on the optimistic assumption that during problem solving, the brain could process unfamiliar, looked up, or calculator-supplied information as effectively as it processed information previously well-memorized.

Unfortunately, by about 2010, cognitive research had found that this assumption was mistaken. Because of evolved characteristics of the "working memory" where the human brain solves problems, as it solves the types of problems we assign in pre-graduate-level physical science courses, the brain applies well-memorized information with relative ease, but is stringently limited when applying facts, relationships, and procedures that have not previously been made quickly recallable from the brain's long-term memory.

One of the implications of this finding is that during problem solving of even minimal complexity, reasoning is likely to be unsuccessful unless it involves primarily the application of previously well-memorized facts and algorithms.

This is a paradigm shift: A change in our fundamental scientific understanding of how students solve problems and how we can best advise them to learn to solve problems. The finding of stringent, quantified working-memory limits means that some of the reform initiatives proposed and adopted in science education in the past two decades will likely need to be re-thought. And, as observers including Max Planck and Thomas Kuhn have noted, even among scientists, accepting a paradigm change can be difficult.[10]

However, cognitive research has also found many of the "active learning" strategies in recent reform proposals, when timed correctly, can promote both interest and long-lasting learning.

This paper will analyze recent findings of cognitive research. Where possible, we will cite research summaries written for educators by leaders in cognitive science. Those summaries will limit technical terminology but include extensive references to peer-reviewed studies. Our hope is to assist instructors in locating knowledge that can improve the alignment of instruction with cognitive science while retaining the best features of recent reform initiatives.

**SCOPE**

In this paper, many findings cited will be conclusions regarding how the brain solves problems that for a time were debated among cognitive experts[11] but have not been substantially contested among cognitive experts for the past ten years. By widely accepted definitions, such findings have become consensus science.[10]

Readers are invited, however, to fact-check how long the assertions we cite have been uncontested among experts who study the brain. At points readers believe our interpretation of how this new science should be applied is arguable, challenge is also invited.

Except where noted, the scope of our topic will be limited to the following questions: When solving well-structured problems in quantitative chemistry and physics courses for science majors, what are the measured strengths and limitations of the brain's working memory? What are the implications of those characteristics? What strategies work around those limits?

**AN OVERVIEW OF LEARNING**

In the following sections, the terms *learning* and *memorization* and *problem solving* will often be used interchangeably. Why?

"Problem solving" in the context of this article would include solving an end-of-chapter textbook problem or answering a clicker question during lecture. It would also include reading a paragraph, looking at a diagram, or encountering a new equation symbol and asking: What does this mean?

A key finding of cognitive research (explored in detail below) is that when the brain solves a problem, it relies almost entirely on recall of stored records of experience (memories) from previous problem solving. When a problem is solved with perceived success, the brain changes physiologically in ways that store memories of how the problem was solved. As memories are applied to solve new problems, stored memories either grow new wired connections to other stored knowledge or improve existing connections. The knowledge and connections enable recall of memories needed to solve new problems.

Mathematics and the physical sciences are composed of precise elements of knowledge not often encountered in our daily experience. To move that type of knowledge into memory, the brain requires repeated effort at recall: memorization. As science instructors in introductory courses, this suggests that one of our primary roles is to identify fundamentals that are the foundation for a new scientific topic and to encourage students to move those fundamentals into recallable memory. Next, by assigning increasingly complex problems, as students solve correctly, the brain wires connections among memories that enable higher-level problem solving.

To summarize: Learning physics and chemistry requires that we store new knowledge in memory, then apply those memories to solve problems.

**TYPES OF PROBLEMS**

Cognitive science divides problems into two types. **Well-structured problems** have a specific goal, initial state, constraints, and precise right answers that can be found by structured procedures. Mathematic and scientific calculations are one example. Cognitive experts view mathematics, chemistry, and physics as overlapping subsets of the language of words and symbols adopted by science to describe the physical universe. Within that language, the brain solves problems in nearly all respects by the same rules.

All other types of problems are termed "ill-structured," including those for which right answers are not known or are debatable.[5,12]

In the United States each year, for each student graduating in physics or chemistry majors combined, about 12 students graduate in biological, health care, or engineering majors.[13] Nearly all students in each of those majors are required to pass a rigorous, quantitative course in both chemistry and physics.

To serve these majors, as widely adopted textbooks reflect, chemistry and physics are designed to teach methods to solve the types of well-structured problems encountered in work across the sciences. For questions in which science knows algorithms that, when applied properly, result in accurate right answers, society expects scientists, engineers, and medical professionals to apply those proven methods rather than less-reliable "heuristics" that tend to involve speculation.

Below we will summarize in bold and then detail findings that, according to cognitive experts, apply when solving well-structured problems.

**INSTINCTIVE AND NON-INSTINCTIVE LEARNING**

**Some learning is automatic; learning mathematics and physical science is difficult.** Into the DNA of all animal species, powerful drives have been programmed by natural selection. Learning is one such drive. Each species has evolved to learn automatically for limited topics that assist with adaptation and survival in primitive and highly competitive environments. For those topics, the brain automatically stores in long-term memory knowledge gained from experience.[14-16]

As one example, our species is driven to learn sophisticated vocalized communication. Creating speech is a task of enormous cognitive complexity, yet human children learn thousands of rules required to speak the dialect of their clan with minimal explicit instruction in those rules. Instinctively, to solve the problem of learning to speak, children store in memory words they hear, subconsciously determine their meaning and rules for their use, then apply the rules as they add the words to their spoken vocabulary.[14]

The human brain also automatically and subconsciously determines and stores complex rules for limited additional topics that assisted in prehistoric survival, including facial recognition, conventions of familial and social relationships, and a basic "folk physics" of how things work.[15,16]

Our brain has weaknesses as well. To focus on survival skills, other learning must be lower priority. As a result, for non-instinctive topics, moving knowledge into memory requires effort. Learning mathematics and physical sciences has additional difficulty in part because, being primarily symbol rather than sound-based, few aspects are learned automatically,[15,16] and in part because their definitions are precise.[4] For example, English-speaking children automatically learn an operational definition of "temperature," but storing verbatim in memory "temperature is a measure of the average kinetic energy of molecules" requires repeated effort at recall.

In the terminology of cognitive science, a type of learning is **primary** when it is instinctive, automatic, and effortless in response to experience. Learning during childhood to speak the local dialect is an example of primary learning.

When learning a topic requires effort, it is termed **secondary**. The central purpose of schooling is secondary learning: to teach citizens to solve problems humans do not learn to solve automatically but modern societies need solved. Learning to solve problems in math and science is secondary: it requires effort and it benefits from structured instruction.

**COGNITIVE ARCHITECTURE**

According to cognitive science, to solve problems of any complexity, the brain must apply previously learned relationships to process problem data. The model science uses to explain learning is based upon the brain's components and their characteristics: what cognitive researchers describe as the brain's "atomic" structure.[17]

Descriptions by cognitive scientists of how the brain solves problems include both observations of neuroscience and models from cognitive psychology that accurately predict experimental outcomes. The following will use terminology and explanations from both disciplines that are employed most often to describe learning.

**Learning involves different types of memory.** Problem solving can be accurately modeled based on the characteristics of long-term memory (LTM), working memory, and activated chunks within LTM.

The brain stores information an individual has learned in **neurons:** specialized cells in the brain's **long-term memory (LTM)**. The brain has tens of billions of neurons that can store small elements of knowledge. Neurons can *grow* thousands of **wires** that connect at **synapses** to wires from neurons storing other elements. Neurons can "fire," sending out a signal with electrical and chemical components to connected neurons or cells.[6,18] Information is stored in overlapping, interconnected networks among neurons. New knowledge becomes stored in LTM as a **memory trace** that changes the neurons, and/or their connections, and/or the synapses at the connections.[19]

This structure gives human LTM an enormous potential capacity, but new knowledge must be stored gradually. As one example, U.S. college graduates on average know the meaning of about 17,000 "root words" learned since infancy at a rate of 2-3 per day.[20]

**Learning begins with attention.** To solve a problem, the brain must first focus its **attention** on the problem's data.[4] Input collected by the senses is then **encoded**: decomposed into small elements the brain can store and process. Those elements move into the brain's **working memory**.[18,19,21]

Working memory is where problems are solved. For multi-step problems, answers are obtained by **information processing**: holding, manipulating, and integrating problem data with previous learning.[22,23]

**Activated chunks** can be described as a third type of memory. If data elements have been found to be related in some way during problem solving, the neuronal networks holding the related elements in LTM tend to grow or strengthen wired connections among the related chunks. Cognitive science defines a **chunk** as related, connected knowledge elements.

In problems assigned in physical science and mathematics courses, the data are typically words, numbers, or symbols previously stored in LTM that have either a few wired connections (stored in LTM as a small chunk) or many connections (stored as part of a larger LTM chunk). **Activated** chunks are neurons in LTM chunks and their connected chunks that have recently fired in response to problem cues.[6,21,24-26]

**The brain also has a "middle-term memory."** After processing in working memory, new chunks and/or their relationships are stored in a "middle-term memory" area in the brain from which they can be recalled for a few days. If recall of the relationship is practiced during that time, it is more likely to be made retrievable from *long*-term memory.

In neuroscience, what happens middle-term is important, but our focus in this article will be limited to: What strategies can our students apply so that information they learn is stored in *long*-term memory, so it is recallable months and years later in courses and careers?

## HOW PROBLEMS ARE SOLVED

Among cognitive experts there is broad agreement on a substantial number of rules that apply when solving well-structured problems.[4-6]

**Cues prompt recall of chunks.** During problem solving, as data enter working memory, the brain uses each chunk as a **cue** to search LTM for matching chunks.[27] If a match is found, the LTM neurons holding the match **activate** ("fire") and send an electrical and chemical signal to other elements to which they are wired (chunked). Depending on the characteristics of the signal, connected chunks may also activate and fire.[6,18,21]

Linked, related chunks that activate are said to be **recallable** into working memory by the cue. In working memory, relationships in activated, recallable chunks *plus* the novel problem data can be **integrated**: processed together to convert problem data step by step to reach the problem goal.[2,18,22,27] An example from chemistry are the steps converting "units to moles to moles to units" in a stoichiometry calculation.

**To learn, solve problems successfully.** How is a new chunk stored in LTM? How do its related elements become linked? Neuroscience educator Efrat Furst explains, "processing in working memory is … information's 'entry ticket'" to LTM.[18]

If a step while solving a problem in working memory is "perceived as useful and successful,"[28] a record of chunks processed at the step is stored in "middle-term" memory. If these chunks are repeatedly processed simultaneously during the next week or two, the relationships tend to become wired into *long*-term memory.[29,30] The (simplified) adage of neuroscience is "Neurons that *fire* together, *wire* together."[31]

However, if similar problems are *not* solved several times over several days, the ability to recall the relationships applied while solving tends to be lost.[18,29,32,33]

An example of wiring LTM is memorizing a word's definition. To begin, an unfamiliar word (a chunk of letters and a chunk of sounds) is defined by terms that state its relationship to chunks of knowledge previously stored in LTM. By repeated practice at recall, as the new word and its meaning are processed together in working memory, the word becomes stored in LTM neuronal networks, and its relationship to the previously stored chunks in its definition becomes wired into LTM.[18,33] By a similar process, other fundamental relationships such as arithmetic facts and scientific equations can be made recallable.[34]

**To learn, start by making fundamentals recallable.** For each new topic, a primary student goal is to learn to solve multiple types of problems. To do so, an initial step is to make the relationships needed most often to solve problems *quickly recallable* from LTM in response to cues encountered in topic problems.[4,27]

**Wire larger chunks by solving problems.** After basic relationships are made recallable from LTM, more complex and conditional relationships can be wired by applying new recallable fundamentals to solve additional problems that include varied cues and contexts. Solving "word problems" is one example, but cognitive processing during demonstrations and laboratory activities can also deepen understanding by wiring sensory elements including sounds, visual images, odors, textures, plus sequences of steps, into chunks.

As chunks are repeatedly processed together, their wired connections become stronger: They activate more easily in similar contexts.[21,24,26] As the elements in chunks and information stored by their wiring increases, their cues and contexts can enable activation of more relationships useful in solving related problems.[18,35]

**Wire a conceptual framework.** As new sub-topics are learned, the expanding chunks in LTM form a **conceptual framework** (also termed a **schema**) for a topic.[36] Chunked elements and the strength of their wired connections provide the brain with an intuitive sense of *which* facts and procedures need to be recalled to solve a problem.[26,28,34] In addition, after a basic schema is wired, additional related information can be stored LTM with less effort.[18,34]

**WORKING MEMORY LIMITS**

A key discovery of recent cognitive research is that solving problems tends to be difficult until relationships needed frequently have been *very* well memorized. Understanding the conditions in which working memory is both strong and is limited can assist in designing instruction that improves student learning.

**Working memory is limited in both duration and capacity.** Working memory can be described as composed of **slots.** During problem solving, these slots must hold the novel (problem specific) chunks of data in a problem. Novel data can include the problem goal, initial problem data, steps to solve, answers found at middle steps, answers to needed relationships that must be "looked up," and final answers.[2,12,24,37-40]

During problem solving, for the novel problem data moved into slots, two limits apply. First, storage in working memory is temporary. Each slot can retain a chunk of novel data for about 30 seconds or less during other processing.[41,42.] Second, at each problem step, if data are described using symbols (such as letters, words, or numbers), working memory in adults usually has only 3-5 slots that can hold a novel data chunk.[39,40]

If data are supplied from multiple senses, such as "dual coded" by providing both visual and auditory descriptions, a few additional slots may be available, but the number of novel slots remains quite limited.[38,43]

**Space for activated chunks is unlimited.** Working memory has the limits described above, but it also has strengths. A key to understanding learning is this rule:

> In working memory, during problem solving, slots are *unlimited* for relationships that can quickly be recalled from LTM. For other types of data chunks, space and holding time are *very* limited.

During cognitive processing, working memory has a strong ability to recall and apply information that is well understood: relationships that have previously been wired into the conceptual frameworks of an individual's LTM.

Kirschner, Sweller, and Clark summarize that working memory is "very limited in duration and capacity" when processing information not well connected in LTM, but "when dealing with previously learned information stored in long-term memory, these limitations disappear."[44]

Why would the brain evolve stringent working memory limits? Sweller notes that the reason for an outcome of natural selection for any trait is speculative, but one factor may be: By limiting the flow of unfamiliar information from the environment into working memory and then via processing to LTM, change in LTM must be gradual. This allows time for careful consideration by the brain of changes in LTM. Care in consolidation of new information may protect the stability of the conceptual frameworks that support our mental well-being (see reference 11, page 136).

**Working memory limits can cause problem solving to fail.** One implication of the rules above is that for a given topic, if a student has moved many fundamental chunks into LTM and wired them to many other relevant chunks, substantial resources are available to solve problems. But what happens if the relevant chunks are few in number or limited in connections?

At steps of a problem, if a cue is supplied but its needed related chunks must be looked up or calculated, novel working memory will need to hold both the cue and its non-recallable related chunk. In a complex problem, the limited slots for novel data may already be full. If this is the case, working memory is said to **overload.** Overload causes a stored data chunk that may be needed at later steps to be "bumped out:" lost from working memory slots. [18,21,22,33] When working memory overloads, Willingham writes, "thinking will likely fail."[45]

For example, when solving a problem, if "phosphate ion" is supplied but its needed formula must be looked up, the formula requires a novel working memory slot. If slots are already full, overload forces out a data chunk, and confusion may result at a later step. In contrast, if an ion name and formula relationship has been memorized as a quickly recallable chunk, the name and formula need only one slot, making overload less likely.

For situations in which the prior wiring of chunks by students is limited, Tracy Alloway and Susan Gathercole advise, (Ref. *38*, p. 134),

> "[T]he capacity of working memory is limited, and the imposition of either excess storage or processing demands in the course of an on-going cognitive activity will lead to catastrophic loss of information from this temporary memory system."

Speed is also important during problem solving. Information well-memorized is recalled quickly. Having to look up an answer, from a table, calculator, or the internet, takes time. Because storage in working memory is quite limited in duration, during a search, other novel problem information being held in slots tends to "time out" and be lost.

**Working memory limits and their impact are not contested.** As previously noted, neuroscience and cognitive psychology explain learning in ways that vary in terminology and detail. However, all current scientific descriptions of problem solving include storage space in working memory that is stringently limited for novel data and non-recallable relationships but is essentially unlimited for relationships quickly recallable.[44,46]

**To summarize:** Working memory (where the brain solves problems) has minimal space for information that has not previously been well memorized. In a given topic, an individual's ability to solve problems depends in large measure on how many relevant facts and procedures can be quickly recalled from previous storage in LTM.

**WORKING AROUND THE LIMITS**

Cognitive scientist Daniel Willingham describes working memory as "the bottleneck of the mind because there is a limited amount of space in working memory."[33] When looking for ways to speed up a process, attention is focused on widening (or working around) the "bottleneck." To guide students in addressing (or circumventing) the bottleneck in learning, cognitive experts advise that we teach students three major strategies: automaticity, chunking, and algorithms.

**Gain automaticity in recall of fundamentals.** For most instructors, the need to memorize relationships as a part of learning physics and chemistry will not be unexpected. What may be surprising is how *thoroughly* science says fundamentals should be memorized. In a 2008 report on learning processes for a U.S. presidential commission, six of the nation's leading cognitive scientists (David Geary, Daniel Berch, Wade Boykin, Susan Embretson, Valerie Reyna, and Robert Siegler) advised (ref. *4*, p. 4-5):

> "[T]here are several ways to improve the functional capacity of working memory. The most central of these is the achievement of automaticity, that is, the fast, implicit, and automatic retrieval of a fact or a procedure from long-term memory."

In cognitive science terminology, *implicit retrieval* is recall that can be described as *intuitive* or *tacit*, often occurs unconsciously, and may not include a conscious ability to explain the reason the particular chunk is recalled.[28]

Geary et al. add:

> "[I]n support of complex problem solving, arithmetic [fundamental] facts and fundamental algorithms should be thoroughly mastered, and indeed, over-learned, rather than merely learned to a moderate degree of proficiency."

Problem solving in mathematics and the physical sciences relies primarily on recall of fundamentals "with automaticity," meaning facts and procedures needed frequently should be *very* well memorized.

**Overlearn by retrieval practice.** Achieving automaticity in recall that lasts for more than a few days requires **overlearning**, defined as "sustained practice, *beyond the point of mastery*" in recall of knowledge from LTM.[33] Overlearning begins with **retrieval practice** activities, such as flashcard use, practice writing from memory tables of relationships, memorizing mnemonics (RoyGBiv), sequence recitation (methane, ethane, …), and answering clicker questions.[33,47,48] Retrieval activities apply the **testing effect**: Recall is strengthened more by testing, including self-testing, than by highlighting or re-reading.[27,49,50]

In Willingham's widely-cited formulation, "memory is the residue of thought."[27,51] We tend to remember what we think about. As an example, during a lecture (or reading assignment) that includes definitions of many new terms, a pause for "clicker questions" on those definitions can help to move new vocabulary from working memory into "middle-term memory." This can free space in working memory for additional new information, assisting with comprehension of speech (or reading) that follows.[18,24] In addition, the recall prompted by a clicker question, if it is an initial step of spaced overlearning, promotes wiring the new vocabulary chunk in LTM.

**Overlearning should be spaced.** To promote wiring facts, procedures, and their linkages in *long*-term memory, overlearning should be repeated over several days and periodically thereafter. This **spaced overlearning** via **spaced retrieval practice** is termed **distributed practice.**[29,32,48,52]

Practicing recall several times in one day (termed **massed practice** or **cramming**) can promote success on a test, but if crammed knowledge is needed again after a few weeks, it tends to require a repeat of intensive study. In contrast, after spaced overlearning, forgetting may occur, but when needed later, quick recall can generally be restored with far less re-study.[52] Cognitive scientists call this finding "savings in relearning."[32]

During careers, work may change, but facts and procedures studied by spaced overlearning tend to be recallable, with occasional "refreshing of memory," for a lifetime.[32,33,53]

**Wire improved chunks.** As students overlearn new relationships, chunks of knowledge are stored in LTM and are wired to related chunks. As those chunks are applied to solve varied problems, additional chunks become connected in LTM and their relationships become more clearly defined. These larger chunks promote improved recall of relevant problem-solving relationships.[23,24]

**Learn effective algorithms.** An **algorithm** (also called a **procedure)** is a "recipe" that solves a type of complex problem in a sequence of steps. Examples include sequences remembered by mnemonics, the standard algorithms of arithmetic and algebra, and the steps of a worked example.

Complex problems can have many chunks of novel data. Successful algorithms are tested, proven sequences that, for a given type of problem, process data in steps that avoid overloading working memory at any step.[4,34,45] Most algorithms are designed to process data by quickly applying well-memorized relationships, circumventing the duration and capacity limits of working memory. To avoid overloading working memory, the steps of algorithms and the contexts in which they apply should be recallable with automaticity, either implicitly or explicitly, from LTM.[4]

Gaining automaticity in the recall of appropriate facts and procedures requires time and effort. John R. Anderson, a leader in research on cognitive processing, advises (ref. *25*, p 359)

> "[A]cquiring competence is very much a labor-intensive business in which one must acquire one-by-one all the knowledge components.… We need to recognize and respect the effort that goes into acquiring competence."

For students, learning science is hard work, but it can be less difficult if we teach strategies recommended by science to work around working memory limits.

**Algorithms are required for all but simple problems.** What does working memory overload feel like? How do algorithms work around limits? Let's perform an experiment.

Try multiplying 84 times 78 "in your head." No fingers or pencil. Try for at least two minutes, *then* write down what you experienced.

Now try 84 times 78 again, this time using pencil and paper.

A standard algorithm you likely memorized long ago uses a pencil instead of working memory to store middle-step digits. Compared to "in your head," how well did using a pencil work?

When trying to reason without the pencil, did storing the middle-step digits cause other needed data to be bumped out or timed out of your working memory?

Two digits times two digits is relatively simple. Most general chemistry problems, of necessity to prepare for future science careers, are not simple. If instructors cannot solve relatively simple problems without reliance on an algorithm, can we expect our students to do so?

Chemists and physicists understand the concepts of arithmetic, but when reasoning to solve a multi-step problem without an algorithm in a domain in which you are not an "expert," the confusion you experienced will likely result.[9,31,32,45] Repeated failure is discouraging. Science-based strategies are likely to lead to success, which may be motivational for prospective science majors.

## IMPLICATIONS FOR INSTRUCTION

With the arrival of the information age, memorization of facts and procedures was widely presumed to have become less important. As a result, a number of recent reforms in physics and chemistry education have been proposed and, in some cases, adopted that assumed students could solve problems by reasoning that relies on looked up information, big ideas, explicit conceptual understanding, inquiry processes, and/or "thinking like a scientist."

Unfortunately, since 2010, the experts who study how the brain works have been in agreement that in fields we are not experts, solving complex problems without applying memorized facts and algorithms is simply not something the brain can manage.

In some cases, cognitive scientists have addressed directly why some of the post-1990 reforms proposed in chemistry and physics education may need to be re-examined.

**In topics learned automatically, you can solve problems by reasoning, but mathematics, physics, and chemistry require algorithms.** Because the human brain automatically stores and wires large quantities of social and survival-skill knowledge, we all become expert in many daily activities. As a result, it may seem that to solve problems, we can reason like an expert generally. But cognitive studies show that for the not-simple but well-structured problems that are the focus of assignments in pre-graduate-school physical science courses, student reasoning without applying algorithms is not likely to reliably lead to correct answers.[24,54]

When students seem to solve complex problems with reasoning but without a recalled procedure, research nearly always careful research finds students are recalling, consciously or subconsciously, a similar procedure that has worked in the past. As Willingham summarizes, careful studies nearly always find "understanding is remembering in disguise."[55] Anderson summarizes, "one fundamentally learns to solve problems by mimicking examples of solutions."[25]

**Learn to apply facts and algorithms "with automaticity."** Some proposals for curriculum reform disparage algorithmic problem solving. Cognitive studies have established that, to the contrary, applying memorized algorithms "with automaticity" is necessary to solve complex problems – to avoid working memory overload.

Richard Clark notes that automated knowledge, which is often applied unconsciously, "allows us to circumvent limits on conscious thinking …. while leaving working memory space to process the novel components of tasks."[28]

Being able to apply facts and procedures quickly and accurately, without conscious thought, is a goal of mastering any complex task. It requires substantial practice.[25,28]

**Students cannot solve by "thinking like a scientist."** Experts can often solve complex problems by reasoning without an algorithm, but students cannot. On the claim that students can be taught to "think like a scientist" or other expert, Willingham writes:

> "[A] flawed assumption underlies the logic, namely that students are capable of doing what scientists or historians do…. No one thinks like a scientist … without a great deal of training…. Real scientists are experts…. [Y]ears of practice make a qualitative, not quantitative, difference in the way they think…."[56]

Kirschner and Hendrick advise, "Beginners aren't 'little' experts."[36]

In a scientific discipline, constructing the vast conceptual framework of an expert generally requires about 10 years of increasingly challenging problem solving.[33]

**Learn concepts, but store fundamentals first.** Concepts are fundamental principles that categorize and explain knowledge, often within a hierarchical structure. Potential energy, conservation of energy, and conservation of matter are concepts with many topic subcomponents. Willingham writes, "conceptual knowledge refers to an understanding of meaning, … understanding *why*" something is true.[34]

To learn a concept, at least some subcomponents of concept knowledge must be quickly recallable first. Willingham advises, "a teacher cannot pour concepts directly into students' heads. Rather, new concepts must build on something students already know."[34]

**Use concepts to solve simple problems, but algorithms for complex**. Teaching concepts helps the brain organize the neuronal networks of its structural frameworks. Applying concepts to solve simple problems helps to wire concepts into LTM.

However, students who aspire to work in science, medical professions, or engineering are also expected to solve complex but well-structured problems. In such cases, due to working memory limits, cognitive science is emphatic: To reliably solve a complex problem, students must apply a well-memorized algorithm.[4,25,55]

Conceptual reasoning for simple problems, but algorithms for complex, has traditionally been how physics and chemistry were taught and learned. What cognitive science has added is an explanation of why, to solve complex problems, applying an algorithm is necessary.

**Research and test effective algorithms.** For a given topic, which algorithms should be taught? Those the instructor has found by research to most often result in right answers. Such algorithms will generally be incremental, highly-structured, easy to remember, relatively quick, widely applicable, have fundamental relationships at their core, and rely on previously well-memorized relationships to circumvent working memory overload.[4,25,34,56]

**Non-majors need implicit, not explicit, understanding.** Non-experts (students) can learn to solve complex problems. To pass chemistry and physics courses, they must. But cognitive studies have established that during problem-solving, the "non-expert" brain bases its decisions on what to do, and what to recall when, on implicit, intuitive, tacit, often unconscious conceptual understanding. In most human activities, the brain acts based on subconscious understanding of what to correctly do without explicitly being able to explain why.[4,25,28]

For example, even without schooling, children above age 3 nearly always correctly apply the complex rules for morphology, syntax, and semantics when they speak their primary language. Yet even as adults, unless we are majors in linguistics, most of us can state very few of the rules that govern our speech.[14,28]

Another example: Most science majors (and their instructors) can use math as a tool to solve scientific problems, but rarely need to recite or explain the rules of mathematics during problem solving. Instructors can solve " $3x + 2 = 29$ , $x = ?$ " in seconds, but in doing so, do we articulate *why* we subtracted before we divided?

No. During initial learning of procedures, the why of steps should be explained. But the long-term goal for those seeking to apply rather than explain mathematics is to automate procedures: To do the right thing *without* having to think about why.[28]

To solve calculations in science, our students do not need to explain the math of each step as they go, and should not. Doing so would tend to cause "timing out" losses from working memory, leading to confusion.

By graduation, we want *majors* in a discipline to be able to explicitly explain why and how their discipline works. However, the vast majority of students in introductory physics and chemistry aspire to major in biological sciences or engineering. For this substantial majority, the substantial additional time required to learn to explain explicit understanding of physics or chemistry must be weighed against the time this would take away from practice solving varied problems. The brain requires practice to gain the implicit understanding needed to choose and recall effective algorithms.

**Guided inquiry?** *After* **learning fundamentals.** Cognitive scientist Daniel Schwartz and colleagues have persuasively argued that guided inquiry activities which "take a few minutes rather than days" can "set the stage" for learning a new topic.[57] But scientific studies also find that longer guided inquiry, at the start of a topic, may result in misconceptions that are difficult to unlearn.[42,58] Cognitive scientist Barak Rosenshine notes scientific studies of effective instruction have consistently found:

> "The most effective teachers … went to hands-on activities, but always after, not before, the basic material was learned."[58]

Cognitive studies tell us that in physical science courses, substantial time and effort must be devoted to overlearning fundamental facts and procedures. Anderson, Reder, and Simon cite the need "to find tasks that provide practice while at the same time sustaining interest."[59] Many of the activities in recent science reform curricula are quite interesting, and cognitive science says can be effective in building frameworks—if conducted "always after" fundamentals are quickly recallable.

**WORK-AROUNDS THAT WORK**

In teaching physics and chemistry, what additional strategies do cognitive experts recommend?

**To limit memorization, instructors should identify fundamentals.** Changing the brain requires biochemical resources. Because the change in LTM wiring that can be constructed physiologically each day is limited, a key question for students is "what is most important to overlearn?" Cognitive experts recommend instructors both identify, and then ask students to overlearn, "core skills and knowledge that will be used again and again."[33]

Reference textbooks cover content in great detail. If a topic introduction by an instructor identifies knowledge essential to construct a basic conceptual framework for the topic, subsequent learning should progress more quickly. Furst summarizes, "Time is better spent at teaching the basics then trying to teach the new without it."[18]

**Overlearn vocabulary early.** For students in introductory science courses, pico-, phosphate, potential energy, proton, photon, P, *P*, Pb, *pH, pKa,* and kPa are the vocabulary of an unfamiliar foreign language. When listening to lecture, reading the text, or solving problems, if multiple words, abbreviations, and symbols are encountered with meanings that cannot be automatically recalled, working memory tends to overload and confusion tends to result.[4,34]

During extended activities, if the student focus is, "what do those new words and symbols mean?" definitions will gradually be learned. However, additional time will need to be spent on additional learning activities to store more complex relationships and context cues necessary to construct a schema for the topic.[33] For student taking multiple courses, time is limited.

Successful problem solving, needed to promote LTM storage and wiring, is more likely if fundamentals are made initially recallable first.

The foundation for new learning is what the student already knows.[18,24,60] Willingham advises,

> "Knowledge is not only cumulative, it grows exponentially. Those with a rich base of factual knowledge find it easier to learn more—the rich get richer."[24]

**Gradually construct meaning.** After fundamental relationships have been made initially recallable, during problem solving, more slots in working memory tend to be available for context cues. When processing includes contextual cues, deeper understanding of the meaning of chunks is wired. More meaning helps the brain judge which facts and procedures to recall for different types of problems.[21,27]

Willingham suggests that instructors

> "[E]xplain to students that automaticity in facts is important because it frees their minds to think about concepts."[34]

**Pre-requisites must be overlearned.** If pre-requisite fundamentals are not recallable, "refreshing memory" should be practiced first.[18] The primary prerequisite for introductory chemistry and physics is quick recall of 10-12 years of arithmetic and algebra. Without this automaticity, working memory tends to overload or time out during scientific calculations. In addition, if students can follow an explanation of a quantitative concept -- such as kinetic energy -- using mental math, slots in working memory are more likely to be available to process conceptual linkages.[34]

Unfortunately, in the U.S., science's discovery of the importance of automaticity has been slow to disseminate in both math and the sciences. As a result, for many current U.S. students, a part of K-12 schooling has included years in curricula that de-emphasized mental math.[37] To improve success in chemistry and physics, review to achieve computational automaticity may be advised.[8,61-65]

**TO SUMMARIZE: WHAT IS LEARNING?**

The purpose of study – or any learning activity -- is to store new knowledge in long-term memory. New knowledge is stored when it is first made recallable, is then used to successfully solve problems, and practice at recall is repeated over time. If an activity does not lead to improved information storage in LTM, nothing long-term has been learned.[18,23]

**ADDITIONAL STRATEGIES FOR LEARNING**

Solving problems in physics and chemistry requires much more than initial memorization. Additional reading is suggested on "interleaved" problem sets,[53] elaboration,[49] dual coding,[43] and the use of demonstrations and inquiry to build frameworks.[42,58] But cognitive experts agree that to begin to learn a new topic, students must achieve quick, automatic recall of fundamentals, and spaced overlearning of those fundamentals should follow. As instructors, can we apply these discoveries to help our students speed, organize, and retain their learning?

**EXPERIMENTS FOR EDUCATORS**

Below are some ideas for formal or informal experiments in the application of cognitive science in courses prior to and including college chemistry and physics. The intent is that modifications of instructional designs be gradual, tested, adapted, and tweaked.  Every class has unique students, goals, and resources.

**Experiment with student preparation for lecture.**  For each new topic, some new components are complex, context dependent, and conditional.  They benefit from written, graphic, and auditory explanation.  But some relationships, such as vocabulary definitions and key equations, are short, precise, and the same across contexts.   These can be learned before a first-day lecture if definitions are posted and a first-day quiz is set.

Might learning well-defined fundamentals before lecture speed and improve learning?  Experiments exploring options might include:

- A day or two before the start of a new topic, in a flashcard or table format, supply 12-20 new relationships essential for solving topic problems.  Announce a 5-minute quiz on their content on topic day one.

- As an alternative, base the first-day quiz on a one-page posted explanation of "basics with some context," asking students to design their own flashcards from the content.

- To motivate on-time preparation, a quiz must count to some extent, but once the value is established, it may suffice to mix "going over" with grading the quiz.

- To promote spaced overlearning, test a rule that on "preparation for lecture" quizzes, 50% of questions will be from previously assigned "fundamentals to overlearn."

**Activate prior learning.**  To begin mid-topic lectures, ask students to solve a problem with content from a previous class but needed today.[58]  Neuron activation from a prior problem lingers and assists with lecture comprehension.[18]

**Contrive "mental math" problems.**  Consider asking students to "estimate, then calculate."  Then, after practice in mental math, design some chapter tests to be "no calculator" (see references *7* and *65*).  If students cannot estimate to check a calculator answer, can we say they understand a quantitative science?

When demonstrating or assigning calculations, contriving "whole number" data that can be solved by mental math leaves slots open for context cues.[34]

**Help students learn to study.**  Consider assigning the brief article at reference *64* describing how students can help themselves learn in higher education.  If quiz questions on the article are promised, attention to content may be enhanced.

**Find the best algorithms.**  Memorized algorithms are required for complex problems, but often many algorithms work.  Which work best?  The question merits research.

**Address prerequisite mathematics.**  Faculty in both Texas and Minnesota have posted short, editable quizzes on the automaticity in mathematics needed for introductory

chemistry and physics (see references *7* and *62*). Given at the start of a course, these assessments may identify topics for which "refreshing of memory" is advised.  Also found may be students who may benefit from 1-2 credits of "math for the sciences" prior to or concurrent with rigorous physical science courses.

**Measure your progress.**  Measurement of results requires comparison to a widely-accepted standard -- such as an examination with national norms.  Quantification of increased departmental success in gateway courses for science and engineering majors may assist department leaders who work to increase resources for instruction.

**Explore additional perspectives.**  For more summaries of the science of learning, see references *58* (at page 19), *60, 18*, and *42*.  For additional detail, see references *35, 36,* and *67*.

**CONCLUSION**

What science has found about memorization may be disappointing, but as scientists, we have no choice but to respect the domain expertise of scientists who study how the brain works.  When young people come to us for help in learning, we are obligated to follow scientific best practices.

Teaching science's strategies of chunking, automaticity, and algorithms can help students work around the working memory bottleneck.  Fundamentals can be mastered by spaced overlearning.  Are such strategies a "drill and practice" approach to learning science?  In part they are, and drill and practice has been a part of what worked for generations who succeeded in gateway science courses.  But in part these new strategies are not the same.

Cognitive science suggests that in a new topic, if instructors identify fundamental facts and procedures, and ask students to practice recall *before* starting to solve problems, needed chunks will wire more quickly.

If instructors assign practice problems in a variety of contexts, and draw attention to the meaning of problem solutions, frameworks will wire more quickly.  Timed appropriately, demonstrations and inquiry problems can move distinctive cues into memory.  By determining and helping students learn the most efficient algorithms, instructors can promote increased student success in courses that are gateways to science majors.

Cognitive science's verification of what works, based on working memory's limitations and strengths, opens extensive opportunities for formal and informal research by educators.  Understanding and applying new science, we can design more effective instruction.

Cognitive science is inherently our science: the science of learning, central science in the work we do.

**AUTHOR INFORMATION**

Corresponding Authors

JudithAnn R. Hartman   *E-mail:  judithann.hartman@gmail.com

Eric A. Nelson    *E-mail:  RNelson696@gmail.com

**ORCID**

JudithAnn R. Hartman:   0000-0003-0186-9405

Eric A. Nelson: 0000-0003-2909-8892

**NOTES**

The authors declare the following competing financial interest(s): Eric Nelson has co-authored textbooks in chemistry.

**REFERENCES**

1. Tang, H.; Pienta, N.  Eye-Tracking Study of Complexity in Gas Law Problems. *J. Chem. Educ.* **2012,** *89* (8), 988-994; DOI: 10.1021/ed200644k.

2. Delazer, M.; Domahs, F., Bartha, L.; Brenneis, C.; Locky, A.; Trieb, T. The Acquisition of Arithmetic Knowledge—An fMRI Study. *Cortex* **2004,** *40,* 166–167.

3. Dhond, R.P.; Marinkovic, K.; Dale, A.M.; Witzel, T.; Halgren, E. Spatiotemporal Maps of Past-Tense Verb Inflection. *Neuroimage* **2003,** *19* (1), 91-100.

4. Geary, D. C.; Boykin, A. W., Embretson, S.; Reyna, V.; Siegler, R.; Berch, D. B.; Graban, J. The Report of the Task Group on Learning Processes; U.S. Department of Education: Washington, DC, 2008. pp 4-2 to 4-6. *http://www2.ed.gov/about/bdscomm/list/mathpanel/report/learning-processes.pdf*  (accessed Nov. 2020). This report contributed to a subsequent work: National Mathematics Advisory Panel. *Foundations for Success: The Final Report of the National Mathematics Advisory Panel*; U.S. Department of Education: Washington, DC, 2008.

5. Spiro, R. J.; DeSchryver, M.  Constructivism.  In *Constructivist Instruction: Success or Failure?* Tobias S., Duffy T., Eds.; Routledge: New York, 2009; pp 106-123.

6. Seidenberg M. *Language at the Speed of Sight: How We Read, Why So Many Cannot, and What Can Be Done About It.*  Basic Books: New York, 2017; pp 138-142.

7. Leopold, D. G. ConfChem Conference on Mathematics in Undergraduate Chemistry Instruction: Strengthening Students' Math Fluencies through Calculator-Free Chemistry Calculations. *J. Chem. Educ.* **2018**, *95* (8), 1432-1433;  DOI: 10.1021/acs.jchemed.8b00113.

8.  Craig, P. R. ConfChem Conference on Mathematics in Undergraduate Chemistry Instruction: Building Student Confidence with Chemistry Computation. *J. Chem. Educ.* **2018**, *95* (8), 1434-1435; DOI: 10.1021/acs.jchemed.8b00091.

9.  Hartman, J. R.; Dahm, D. J.; Nelson, E. A.  ConfChem Conference On Flipped Classroom: Time-Saving Resources Aligned With Cognitive Science. *J. Chem. Educ.* **2015**, *92* (9), 1568-1569; DOI: 10.1021/ed5009156.

10. Kuhn, T. S.: *The Structure of Scientific Revolutions*. University of Chicago Press: Chicago, 1962; pp 143-158.

11. *Constructivist Instruction:  Success or Failure?*  Tobias, S.; Duffy, T. M., Eds.; Routledge: New York, 2009.

12. Simon, H. A. The Structure of Ill-Structured Problems. In *Models of Discovery.* Springer: Dordrecht, 1977; pp 304-325.

13. Trapani, J.; Hale, K. Higher Education in Science and Engineering. *Science & Engineering Indicators 2020*. NSB-2019-7. National Science Foundation. 2019**,** Table S2-6.

14. Pinker, S. A. *The Language Instinct: How the Mind Creates Language.* HarperCollins: New York, 2007.

15. Geary, D. C. Principles of Evolutionary Educational Psychology. *Learning and Individual Differences* **2002,** *12*, 317-345.

16. Geary, D.C.; Berch, D.B. Evolution and Children's Cognitive and Academic Development. In *Evolutionary Perspectives on Child Development and Education* Springer: Cham, 2016; pp. 217-249.

17. Anderson, J.; Lebiere, C., (Eds.): The Atomic Components of Thought.  Erlbaum Publishers: Mahwah, NJ 1998

18. Furst, E.  Learning In the Brain.  *https://sites.google.com/view/efratfurst/learning-in-the-brain* . (accessed Nov. 2020).

19. Tonegawa, S.; Liu, X; Ramirez, S.; Redondo, R. Memory Engram Cells Have Come of Age. *Neuron* **2015**, *87* (5), 918-31.

20. Biemiller, A. Teaching Vocabulary. *Am. Educ.* **2001**, *25*, 24-28.

21. Furst, E.  Understanding Understanding. *https://sites.google.com/view/efratfurst/understanding-understanding* (accessed Nov. 2020).

22. Alloway, T. P.; Gathercole, S. E. Working Memory and Classroom Learning. *Dyslexia Rev*. **2004** (*15*), 4-9.

23. Clark, R. E.; Sweller, J.; Kirschner, P. A. Putting Students on the Path to Learning: The Case for Fully Guided Instruction. *Am. Educ.* **2012**, *36* (1), 6-11.

24. Willingham, D. T. How Knowledge Helps. *Am. Educ.* **2006,** *30* (1), 30-37.

25. Anderson, J. R. ACT: A Simple Theory of Complex Cognition. *American Psychologist*. **1996,** *51* (4) 355-365.

26. Anderson, J. R.; Bothell, D.; Byrne, M. D.; Douglass, S.; Lebiere, C.; Qin, Y.. An Integrated Theory of the Mind. *Psychological Review*, **2004**, *111* (4), 1036-1060.

27. Willingham, D. T. What Will Improve a Student's Memory? *Am. Educ.* **2008**, *32* (4), 17-25.

28. Clark, R. E. Not Knowing What We Don't Know: Reframing the Importance of Automated Knowledge for Educational Research. In *Avoiding Simplicity, Confronting Complexity*. Clareboutand G., Elen, J., Eds.; Sense Publishers: Rotterdam, 2006, pp 3-14.

29. Trafton, A. Neuroscientists Identify Brain Circuit Necessary for Memory Formation. *http://news.mit.edu/2017/neuroscientists-identify-brain-circuit-necessary-memory-formation-0406* (accessed Nov. 2020).

30. Furst, E. Reconsolidation. *https://sites.google.com/view/efratfurst/reconsolidation* . (accessed Nov. 2020).

31. Hebb, D.: *The Organization of Behavior*. Wiley: New York, 1949.

32. Willingham, D. T. Do Students Remember What They Learn In School? *Am. Educ.* **2015**, *39* (3), 33-38.

33. Willingham, D. T. Practice Makes Perfect—But Only If You Practice Beyond the Point of Perfection. *Am. Educ.* **2004**, *28* (1), 31-33.

34. Willingham, D. T. Is It True That Some People Just Can't Do Math? *Am. Educ.* **2009**, *33* (4), 14-19.

35. Willingham, D. T. Why Don't Students Like School: A Cognitive Scientist Answers Questions About How the Mind Works and What It Means for the Classroom. Wiley: New York, 2009, pp 101-104.

36. Kirschner, P. A.; Hendrick, C. *How Learning Happens: Seminal Works in Educational Psychology and What They Mean in Practice*. Routledge: London, 2020. pp 4-12.

37. Luck S. J.; Vogel E. K. Visual Working Memory Capacity: from Psychophysics and Neurobiology to Individual Differences. *Trends in Cognitive Sciences.* **2013**, *17* (8),391-400

38. Alloway, T. P.; Gathercole, S. E. How Does Working Memory Work In the Classroom? *Educational Research and Reviews* **2006**, *1* (4), 134-139.
22

39. Cowan, N. The Magical Number 4 In Short-Term Memory: A Reconsideration of Mental Storage Capacity. *Behav. and Brain Sciences* **2001**, *24*, 87-185.

40. Cowan, N. The Magical Mystery Four: How Is Working Memory Capacity Limited, and Why? *Curr. Dir. in Psych. Sci.* **2010**, *19* (1), 51-57.

41. Peterson, L.; Peterson, M. Short-Term Retention of Individual Verbal Items. *J. of Experimental Psychol.* **1959**, *58*, 193-198.

42. Clark, R. E.; Sweller, J.; Kirschner, P. A. Putting Students on the Path to Learning: The Case for Fully Guided Instruction. *Am. Educ.* **2012**, *36* (1), 6-11.

43. Kirschner, P A.; Hendrick, C. *How Learning Happens: Seminal Works in Educational Psychology and What They Mean in Practice*. Routledge: London, 2020. pp 41-49.

44. Kirschner, P. A.; Sweller, J.; Clark, R. E. Why Minimal Guidance During Instruction Does Not Work: An Analysis of the Failure Of Constructivist, Discovery, Problem Based, Experiential and Inquiry-Based Teaching. *Educ. Psychologist* **2006**, *41*, 75-86.

45. Willingham, D. T. Why Don't Students Like School? *Am. Educ.* **2009**, *33* (2), 4-13.

46. Cowan N. The Many Faces of Working Memory and Short-Term Storage. *Psychonomic Bulletin & Review*. **2017**, *24*(4), 1158-70.

47. Agarwal, P. K.; Roediger, H. L.; McDaniel, M. A.; McDermont, K. B. *How to Use Retrieval Practice to Improve Learning*. Washington University in St. Louis: Institute of Education Sciences, 2013. *http://pdf.retrievalpractice.org/RetrievalPracticeGuide.pdf* (accessed Nov. 2020).

48. Carpenter, S. K.; Agarwal, P. K. *How to Use Spaced Retrieval Practice to Boost Learning.* Iowa State University:Ames, IA, 2019. *http://pdf.retrievalpractice.org/SpacingGuide.pdf* (accessed Nov. 2020).

49. Dunlosky, J. Strengthening the Student Toolbox: Study Strategies to Boost Learning. *Am. Educ.* **2013**, *37* (3), 12-21.

50. Brown, P.; Roediger; H. L., McDaniel, M. A. *Make It Stick: The Science of Successful Learning*. Harvard University Press: Cambridge, MA, 2014.

51. Willingham, D. T. Why Don't Students Like School: A Cognitive Scientist Answers Questions About How the Mind Works and What It Means for the Classroom. Wiley: New York, 2009, pp 41-66.

52. Willingham, D. T. Allocating Student Study Time: Massed Versus Distributed Practice. *Am. Educ.* **2002**, *26* (2), 37-39.

53. Bjork, R. A.; Bjork, E. L. Forgetting As the Friend of Learning: Implications for Teaching and Self-Regulated Learning. *Adv. Physiol. Educ.* **2019**, *43*, 164-167.

54. Sweller, J.; Clark, R. E.; Kirschner, P. A. Teaching General Problem-Solving Skills Is Not A Substitute for, Or A Viable Addition To, Teaching Mathematics. *Notices of the Am. Math. Soc.,* **2010**, *57* (10), 1303-1304.

55. Willingham, D. T. Why Don't Students Like School: A Cognitive Scientist Answers Questions About How the Mind Works and What It Means for the Classroom. Wiley: New York, 2009, pp 68-72.

56. Willingham, D. T. Why Don't Students Like School: A Cognitive Scientist Answers Questions About How the Mind Works and What It Means for the Classroom. Wiley: New York, 2009, pp 97-98.

57. Schwartz, D. L.; Chase, C. C.; Oppezzo, M. A.; Chin, D. B. Practicing Versus Inventing With Contrasting Cases. *Journal of Educational Psychology* **2011,** *103* (4), 759-777.

58. Rosenshine, B. Principles of Instruction: Research-Based Strategies That All Teachers Should Know. *Am. Educ.* **2012**, *36* (1), 12-19.

59. Anderson, J. R.; Reder, L. M.; Simon, H. A. Applications and Misapplications of Cognitive Psychology to Mathematics Education. *Texas Educational Review* **2000**, Summer.

60. Deans for Impact. *The Science of Learning*. Deans for Impact: Austin, 2015. *https://deansforimpact.org/wp-content/uploads/2016/12/The_Science_of_Learning.pdf* (accessed Nov. 2020).

61. Nelson, E. A. ConfChem Conference on Mathematics in Undergraduate Chemistry Instruction: Addressing Math Deficits with Cognitive Science. *J. Chem. Educ.* **2018**, *95* (8), 1440-1442; DOI: 10.1021/acs.jchemed.8b00085.

62. Williamson, V. M.; Walker, D. R.; Chuu, E.; Broadway, S.; Mamiya, B.; Powell, C. B.; Shelton, G. R.; Weber, R.; Dabney, A. R.; Mason, D. Impact of Basic Arithmetic Skills On Success In First-Semester General Chemistry. *Chem. Educ. Research and Practice* **2020**, *21*, 51-61; DOI: 10.1039/C9RP00077A.

63. Ranga J. S. ConfChem Conference on Mathematics in Undergraduate Chemistry Instruction: Impact of Quick Review of Math Concepts. *J. Chem. Educ.* **2018**, *95*(8), 1430-1431. DOI: 10.1021/acs.jchemed.8b00070

64. Kilner W. C. ConfChem Conference on Mathematics in Undergraduate Chemistry Instruction: The Chem-Math Project. *J. Chem. Educ.* **2018**, *95* (8), 1436-1437. DOI: 10.1021/acs.jchemed.8b00075

65. Penn, L. S. ConfChem Conference on Mathematics in Undergraduate Chemistry Instruction: Estimation—An Empowering Skill for Students in Chemistry and Chemical Engineering. *J. Chem. Educ.* **2018**, *95* (8), 1426-1427; DOI: 10.1021/acs.jchemed.8b00363.

66. Putnam, A. L.; Sungkhasettee, V. W.; Roediger, H. L. Optimizing Learning in College: Tips from Cognitive Psychology. *Perspect. Psych. Sci.* **2016**, *11* (5), 652-660.

67. Hartman, J. R.; Nelson E. A. "Do We Need To Memorize That?" Or Cognitive Science for Chemists. *Foundations of Chemistry* **2015**, *17* (3), 263-274.

# # # # #